\documentclass[12pt]{article}
\begin{document}

\title{The Simplest Massive Lagrangian Without Higgs Mechanism}
\author{G.Quznetsov \\
quznets@geocities.com}
\date{September 17, 1998}
\maketitle

\begin{abstract}
The simplest massive Lagrangian, invariant for the global SU(2)
transformation, is constructed.
\end{abstract}

I use the notations:

\[
\beta ^1=\left[ 
\begin{array}{cccc}
0 & 1 & 0 & 0 \\ 
1 & 0 & 0 & 0 \\ 
0 & 0 & 0 & -1 \\ 
0 & 0 & -1 & 0
\end{array}
\right] \mbox{, }\beta ^2=\left[ 
\begin{array}{cccc}
0 & -i & 0 & 0 \\ 
i & 0 & 0 & 0 \\ 
0 & 0 & 0 & i \\ 
0 & 0 & -i & 0
\end{array}
\right] \mbox{,} 
\]

\[
\beta ^3=\left[ 
\begin{array}{cccc}
1 & 0 & 0 & 0 \\ 
0 & -1 & 0 & 0 \\ 
0 & 0 & -1 & 0 \\ 
0 & 0 & 0 & 1
\end{array}
\right] \mbox{, }\beta ^4=\left[ 
\begin{array}{cccc}
0 & 0 & i & 0 \\ 
0 & 0 & 0 & i \\ 
-i & 0 & 0 & 0 \\ 
0 & -i & 0 & 0
\end{array}
\right] \mbox{,} 
\]

\[
\gamma ^0=\left[ 
\begin{array}{cccc}
0 & 0 & 1 & 0 \\ 
0 & 0 & 0 & 1 \\ 
1 & 0 & 0 & 0 \\ 
0 & 1 & 0 & 0
\end{array}
\right] \mbox{, }\beta ^0=\left[ 
\begin{array}{cccc}
1 & 0 & 0 & 0 \\ 
0 & 1 & 0 & 0 \\ 
0 & 0 & 1 & 0 \\ 
0 & 0 & 0 & 1
\end{array}
\right] \mbox{.} 
\]

In the {Standard Model} we have got the following entities:

the right electron state vector $e_R$,

the left electron state vector $e_L$,

the electron state vector $e$ ($e=\left[ 
\begin{array}{c}
e_R \\ 
e_L
\end{array}
\right] $),

the left neutrino state vector $\nu _L$,

the zero vector right neutrino $\nu _R$.

the unitary $2\times 2$ matrix $U$ of the isospin transformation.($SU(2)$).

This matrix acts on the vectors of the kind:$\left[ 
\begin{array}{c}
\nu _L \\ 
e_L
\end{array}
\right] $.

Therefore, in this theory: if

\[
U=\left[ 
\begin{array}{cc}
u_{1,1} & u_{1,2} \\ 
u_{2,1} & u_{2,2}
\end{array}
\right] 
\]

then the matrix

\[
\left[ 
\begin{array}{cccc}
1 & 0 & 0 & 0 \\ 
0 & u_{1,1} & 0 & u_{1,2} \\ 
0 & 0 & 1 & 0 \\ 
0 & u_{2,1} & 0 & u_{2,2}
\end{array}
\right] 
\]

operates on the vector

\[
\left[ 
\begin{array}{c}
e_R \\ 
e_L \\ 
\nu _R \\ 
\nu _L
\end{array}
\right]. 
\]

Because $e_R$, $e_L$, $\nu _R$, $\nu _L$ are the two-component vectors then

\[
\left[ 
\begin{array}{c}
e_R \\ 
e_L \\ 
\nu _R \\ 
\nu _L
\end{array}
\right] \mbox{ is }\left[ 
\begin{array}{c}
e_{R1} \\ 
e_{R2} \\ 
e_{L1} \\ 
e_{L2} \\ 
0 \\ 
0 \\ 
\nu _{L1} \\ 
\nu _{L2}
\end{array}
\right] 
\]

and

\[
\left[ 
\begin{array}{cccc}
1 & 0 & 0 & 0 \\ 
0 & u_{1,1} & 0 & u_{1,2} \\ 
0 & 0 & 1 & 0 \\ 
0 & u_{2,1} & 0 & u_{2,2}
\end{array}
\right] \mbox{ is }\underline{U}=\left[ 
\begin{array}{cccccccc}
1 & 0 & 0 & 0 & 0 & 0 & 0 & 0 \\ 
0 & 1 & 0 & 0 & 0 & 0 & 0 & 0 \\ 
0 & 0 & u_{1,1} & 0 & 0 & 0 & u_{1,2} & 0 \\ 
0 & 0 & 0 & u_{1,1} & 0 & 0 & 0 & u_{1,2} \\ 
0 & 0 & 0 & 0 & 1 & 0 & 0 & 0 \\ 
0 & 0 & 0 & 0 & 0 & 1 & 0 & 0 \\ 
0 & 0 & u_{2,1} & 0 & 0 & 0 & u_{2,2} & 0 \\ 
0 & 0 & 0 & u_{2,1} & 0 & 0 & 0 & u_{2,2}
\end{array}
\right] . 
\]

This matrix has got eight orthogonal normalized eigenvectors $s_1$, $s_2$, $%
s_3$, $s_4$, $s_5$, $s_6$, $s_7$, $s_8$ of type:

\[
\underline{s_1}=\left[ 
\begin{array}{c}
1 \\ 
0 \\ 
0 \\ 
0 \\ 
0 \\ 
0 \\ 
0 \\ 
0
\end{array}
\right] \mbox{, }\underline{s_2}=\left[ 
\begin{array}{c}
0 \\ 
1 \\ 
0 \\ 
0 \\ 
0 \\ 
0 \\ 
0 \\ 
0
\end{array}
\right] \mbox{, }\underline{s_3}=\left[ 
\begin{array}{c}
0 \\ 
0 \\ 
\varpi \\ 
0 \\ 
0 \\ 
0 \\ 
\chi \\ 
0
\end{array}
\right] \mbox{, }\underline{s_4}=\left[ 
\begin{array}{c}
0 \\ 
0 \\ 
0 \\ 
\varpi \\ 
0 \\ 
0 \\ 
0 \\ 
\chi
\end{array}
\right] , 
\]

\[
\underline{s_5}=\left[ 
\begin{array}{c}
0 \\ 
0 \\ 
0 \\ 
0 \\ 
1 \\ 
0 \\ 
0 \\ 
0
\end{array}
\right] \mbox{, }\underline{s_6}=\left[ 
\begin{array}{c}
0 \\ 
0 \\ 
0 \\ 
0 \\ 
0 \\ 
1 \\ 
0 \\ 
0
\end{array}
\right] \mbox{, }\underline{s_7}=\left[ 
\begin{array}{c}
0 \\ 
0 \\ 
\chi ^{*} \\ 
0 \\ 
0 \\ 
0 \\ 
-\varpi ^{*} \\ 
0
\end{array}
\right] \mbox{, }\underline{s_8}=\left[ 
\begin{array}{c}
0 \\ 
0 \\ 
0 \\ 
\chi ^{*} \\ 
0 \\ 
0 \\ 
0 \\ 
-\varpi ^{*}
\end{array}
\right] . 
\]

with the corresponding eigenvalues : $1$, $1$, $\exp \left( i\cdot \lambda
\right) $, $\exp \left( i\cdot \lambda \right) $,

$1$,$1$,$\exp \left( -i\cdot \lambda \right) $, $\exp \left( -i\cdot \lambda
\right) $.

These vectors constitute the orthogonal basis in this 8-dimensional space.

If $O$ is zero $4\times 4$ matrix then let $\underline{\gamma ^0}=\left[ 
\begin{array}{cc}
\gamma ^0 & O \\ 
O & \gamma ^0
\end{array}
\right] $ and \underline{$\beta ^4$}$=\left[ 
\begin{array}{cc}
\beta ^4 & O \\ 
O & \beta ^4
\end{array}
\right] $.

The vectors:

\[
\underline{e}=\left[ 
\begin{array}{c}
e_{R1} \\ 
e_{R2} \\ 
e_{L1} \\ 
e_{L2} \\ 
0 \\ 
0 \\ 
0 \\ 
0
\end{array}
\right] \mbox{, }\underline{e_R}=\left[ 
\begin{array}{c}
e_{R1} \\ 
e_{R2} \\ 
0 \\ 
0 \\ 
0 \\ 
0 \\ 
0 \\ 
0
\end{array}
\right] \mbox{, }\underline{e_L}=\left[ 
\begin{array}{c}
0 \\ 
0 \\ 
e_{L1} \\ 
e_{L2} \\ 
0 \\ 
0 \\ 
0 \\ 
0
\end{array}
\right] 
\]

correspond to the state vectors $e$, $e_R$ and $e_L$ resp.

For the vector \underline{$e$} the numbers $k_3$, $k_4$, $k_7$, $k_8$ exist,
for which: 
\[
\underline{e}=(e_{R1}\cdot \underline{s_1}+e_{R2}\cdot \underline{s_2}%
)+(k_3\cdot \underline{s_3}+k_4\cdot \underline{s_4})+(k_7\cdot \underline{%
s_7}+k_8\cdot \underline{s_8})\mbox{.} 
\]

Here

\[
\underline{e_R}=(e_{R1}\cdot \underline{s_1}+e_{R2}\cdot \underline{s_2})%
\mbox{.} 
\]

If

\[
\begin{array}{c}
\underline{e_{La}}=(k_3\cdot \underline{s_3}+k_4\cdot \underline{s_4})%
\mbox{,} \\ 
\underline{e_{Lb}}=(k_7\cdot \underline{s_7}+k_8\cdot \underline{s_8})
\end{array}
\]

then

\[
\begin{array}{c}
\underline{U}\cdot \underline{e_{La}}=\exp \left( i\cdot \lambda \right)
\cdot \underline{e_{La}}\mbox{,} \\ 
\underline{U}\cdot \underline{e_{Lb}}=\exp \left( -i\cdot \lambda \right)
\cdot \underline{e_{Lb}}\mbox{.}
\end{array}
\]

Let for all $k$ ($1\leq k\leq 8$):

\[
\underline{h_k}=\underline{\gamma ^0}\cdot \underline{s_k}\mbox{.} 
\]

The vectors \underline{$h_k$} constitute the orthogonal basis, too. And the
numbers $r_3$, $r_4$, $r_7$, $r_8$ exist, for which:

\[
\underline{e_R}=(r_3\cdot \underline{h_3}+r_4\cdot \underline{h_4}%
)+(r_7\cdot \underline{h_7}+r_8\cdot \underline{h_8})\mbox{.} 
\]

Let

\[
\begin{array}{c}
\underline{e_{Ra}}=(r_3\cdot \underline{h_3}+r_4\cdot \underline{h_4})%
\mbox{,} \\ 
\underline{e_{Rb}}=(r_7\cdot \underline{h_7}+r_8\cdot \underline{h_8})%
\mbox{,} \\ 
\underline{e_a}=\underline{e_{Ra}}+\underline{e_{La}}\mbox{ and }\underline{%
e_b}=\underline{e_{Rb}}+\underline{e_{Lb}}\mbox{.}
\end{array}
\]

Let us denote:

\begin{eqnarray*}
\underline{e_a} &=&R_U^{(+)}\cdot \underline{e}\mbox{, } \\
\underline{e_b} &=&R_U^{(-)}\cdot \underline{e}\mbox{.}
\end{eqnarray*}

$R_U^{(+)}$ and $R_U^{(-)}$ are the linear transformations (the $8\times 8$
complex matrices) in the space of $s_1$, $s_2$, $s_3$, $s_4$, $s_5$, $s_6$, $%
s_7$, $s_8$.

It is obviously that the value of

\[
\left( \left( \underline{e}^{\dagger }\cdot \gamma ^0\cdot R_U^{(+)}\cdot 
\underline{e}\right) ^2+\left( \underline{e}^{\dagger }\cdot \beta ^4\cdot
R_U^{(+)}\cdot \underline{e}\right) ^2\right) ^{0.5}+ 
\]

\[
+\left( \left( \underline{e}^{\dagger }\cdot \gamma ^0\cdot R_U^{(-)}\cdot 
\underline{e}\right) ^2+\left( \underline{e}^{\dagger }\cdot \beta ^4\cdot
R_U^{(-)}\cdot \underline{e}\right) ^2\right) ^{0.5} 
\]

does not depend from the choice of the $SU(2)$ matrix $U$.

In this case the Lagrangian:

\[
\begin{array}{c}
L=0.5\cdot i\cdot \left( \left( \partial _\mu \underline{e}\right) \cdot
\beta ^\mu \cdot \underline{e}-\underline{e}^{\dagger }\cdot \beta ^\mu
\cdot \left( \partial _\mu \underline{e}\right) \right) -\  \\ 
-m\cdot \left( \left( \underline{e}^{\dagger }\cdot \gamma ^0\cdot
R^{(+)}\cdot \underline{e}\right) ^2+\left( \underline{e}^{\dagger }\cdot
\beta ^4\cdot R^{(+)}\cdot \underline{e}\right) ^2\right) ^{0.5}- \\ 
-m\cdot \left( \left( \underline{e}^{\dagger }\cdot \gamma ^0\cdot
R^{(-)}\cdot \underline{e}\right) ^2+\left( \underline{e}^{\dagger }\cdot
\beta ^4\cdot R^{(-)}\cdot \underline{e}\right) ^2\right) ^{0.5}\mbox{.}
\end{array}
\]

is invariant for the global $SU(2)$ transformation.

\end{document}